\documentclass[twocolumn,showpacs,superscriptaddress]{revtex4}
\usepackage{graphicx, subfigure}
\usepackage{dcolumn}% Align table columns on decimal point
\usepackage{bm}% bold math
\usepackage{amssymb, amsmath}
\usepackage{times}
\usepackage{epsfig}

\begin{document}

\title{Dissecting the hydrogen bond: a Quantum Monte Carlo approach}

%commentato per footnote
\def\thefootnote{\fnsymbol{ctr}}

%this is for REVTEX STYLE
% 
 \author{Fabio Sterpone $^{1}$,Leonardo Spanu $^{2,3}$,
 Luca Ferraro$^1$, Sandro Sorella$^2$, Leonardo Guidoni $^{4,5}$}
 
 \affiliation{
 $^1$CASPUR, Via dei Tizii 6B, 00185, Roma, Italy\\
 $^2$International School for Advanced Studied (SISSA/ISAS),
 Via Beirut 4, 34014 Trieste, Italy \\
 $^3$ Department of Chemistry, University of California, Davis 95616 USA \\
 $^4$ Dipartimento di Fisica, La Sapienza - Universit\`a di Roma, 
 P.le A. Moro 2, 00185 Roma, Italy  \\
 $^5$ To whom correspondence should be addressed. Email:
 leonardo.guidoni@uniroma1.it 
% $^*$ FS and LS have contributed equally to this work.
 }

%this is for ACS STYLE

% \author{
% Fabio Sterpone\footnote{CASPUR, Via dei Tizii 6B, 00185, Roma, Italy.}
%\footnote{FS and LS have contributed equally to this work.},
% Leonardo Spanu\footnote{International School for Advanced Studied (SISSA/ISAS),
% Via Beirut 4, 34014 Trieste, Italy.}
% \footnote{Current address: Department of
% Chemistry, University of California, Davis, USA.} \footnotemark[2],
%  Luca Ferraro\footnotemark[1],\\
%  Sandro Sorella\footnotemark[2],
%  and Leonardo Guidoni\footnote{Dipartimento di
% Fisica, La Sapienza - Universit\`a di Roma, P.le A. Moro 2, 00185 Roma,
% Italy and NAST Centre – Nanoscience \& Nanotechnology \& Instrumentation,
%Università degli Studi di Roma Tor Vergata, Roma, Italy } 
%\footnote{ To whom correspondence should be addressed. Email:
%  leonardo.guidoni@uniroma1.it, http://bio.phys.uniroma1.it.}
% }

\begin{abstract}
We present a Quantum Monte Carlo study of the dissociation energy 
and the dispersion curve of the water dimer, a prototype of hydrogen 
bonded system. Our calculations 
are based on a wave function which is a modern and fully correlated 
implementation of the Pauling's valence bond idea: 
the Jastrow Antisymmetrised Geminal Power (JAGP) [Casula et al
J.Chem.Phys. 119,6500-6511 (2003)].
With this variational wave function 
we obtain a binding energy of
 $-4.5(0.1)$ kcal/mol, that is only slightly increased to $-4.9(0.1)$ kcal/mol
by using the Lattice Regularized Diffusion Monte Carlo (LRDMC).  
This projection technique 
allows  to improve substantially the correlation energy of a given 
variational guess and indeed, when applied to the JAGP, 
yields a binding energy in fair agreement
 with the value of $-5.0$ kcal/mol reported by experiments and other theoretical
works.
The minimum position, the curvature and the asymptotic behavior of the
dispersion curve are well reproduced both at the variational and LRDMC
level. Moreover, thanks to the simplicity and the accuracy of our 
variational approach,  we are able  
to dissect the various contributions to 
the binding energy of the water dimer in a systematic and 
controlled way. 
This is achieved 
by appropriately switching off determinantal and Jastrow 
variational terms in the JAGP. Within this scheme,
we estimate that the van der Waals  contribution 
to the electron correlation is substantial and amounts to 
 $1.5(0.2)$ kcal/mol, this value being comparable with  
the  intermolecular covalent energy, that we find to be 
 $1.1(0.2)$ kcal/mol.
The present Quantum Monte Carlo approach based on the JAGP wavefunction 
reveals as  a promising tool for the interpretation and 
the  quantitative description
of weakly interacting systems, 
where both dispersive and covalent energy contributions 
play an important role.
\end{abstract}

\date{\today}
\bigskip
\noindent {\bf Keywords: water dimer, valence bonds, 
wavefunction methods, van der Waals interactions, weak interactions}

\maketitle

%=====================================================================
\section{Introduction}
The hydrogen bond is a fundamental intramolecular and 
intermolecular interaction determining 
the properties of a large number of systems from liquids to solids, 
from biological \cite{creighton1993}
to inorganic \cite{steiner2002}.
Hydrogen bond is commonly defined as a local bond in which an hydrogen
atom is attached to an electronegative group (the donor) interacting
with another nearby electronegative group (the acceptor), not
covalently attached.  Dissociation energies cover a range of about two
orders of magnitude, ranging from $-0.2$ to $40$ kcal/mol, the
H-bonding arising from the interplay of different types of
interactions.  Electrostatic forces play the major role in a large
number of hydrogen bonds, although charge transfer effects and van der
Waals (vdW) interactions are always present.

Water, the most studied H-bonding liquid, represents the prototype 
of hydrogen bonded systems. The energetics and directionality
of water hydrogen bond is a key factor for understanding
the anomalous properties of the water phase diagram \cite{franks1972},
the behavior of small water clusters \cite{ludwig2001,xantheas2000,shin2004}
and the role of aqueous environment in a variety of biological systems \cite{ball2008}. 
The dissociation energy of the isolated water 
dimer lies  in the middle of the hydrogen bond dissociation energy scale, 
the most stable configuration being associated with 
a binding energy $D^{exp}_{e} = -5.0$ kcal/mol, as extrapolated by 
experimental data \cite{mas2000}.
The partitioning of this energy in  different contribution terms is still 
subject of vivid debate. At the equilibrium bonding distance, typically 
in a range between $2.5$ and $3.5$ {\AA}, quantum effects become relevants
and a pure electrostatic picture of the interaction is not
fully satisfactory.
The partial covalent nature of the hydrogen bond has been recently 
invoked by  a first analysis of the  
Compton profile on ice $Ih$ \cite{barbiellini1999}.  
However, the interpretation of the experimental data has been questioned
by several authors \cite{ghanty2000,romero2000} and further revised
\cite{barbiellini2002}. The amount of the 
intermolecular covalent contribution, if any, to the binding 
energy is still an open issue. On the other hand, due to
the lack of an unambiguous computational protocol 
it is still not clear how to estimate the van der Waals contribution 
to the hydrogen bonding. The role of 
these interactions may also be at the basis of the current 
drawbacks of empirical forcefields in use for large scale simulations
\cite{cho1997,guillot2002}.

A definition of the intermolecular covalent component 
of the 
hydrogen bonding can be drawn using the intuitive picture of chemical bond 
introduced by Pauling as the superposition 
of Lewis' structures \cite{bratoz1967}. 
In the simple case of hydrogen bonding in a water dimer $(H_2O)_2$, three 
mesomeric Lewis structures may be drawn, one of them describing 
the charge transfer situation $(OH)^- \cdots (H_3O)^+$, 
that confers partial covalent character to the hydrogen bond. 
Within a quantitative Valence Bond representation it would be 
therefore possible to distinguish, in a simple way, the covalent
intermolecular energy contribution by the other interaction 
energy terms.

At the same time, because of the crucial role of electronic correlation,
especially for dispersive interactions,
high quality electronic structure correlated methods
(based on molecular orbital theory) are necessary to a proper 
quantitative description of hydrogen bond.
New classes of Density Functional Theory (DFT) functionals 
have been developed in the past years with the aim to describe 
weakly bound systems avoiding semi empirical approaches.
Nevertheless, the highly non perturbative and non-local character of the vdW 
interactions makes difficult their inclusions in DFT schemes
without the resorting to {\it ad hoc} empirical parameterizations
\cite{vonlilienfeld2004}. 
Looking for an {\it ab initio} method free of empiricism, the 
Quantum Monte Carlo \cite{sorella2007,gurtubay2007}
appears as a possible alternative to other
more standard quantum chemistry methods such as
Configuration Interaction, M\"{o}ller Plesset perturbation theory 
or coupled-cluster (CC).

Recently, a
QMC technique  based on the  resonating valence bond 
wavefunction was introduced in Ref.\cite{casula2003},  and 
further developed later.
This approach represents  a very efficient  implementation of the 
 valence bond Pauling idea, discovered 
by P.W. Anderson in the field of strongly correlated electrons \cite{anderson1987}:
the Jastrow Antisymmetrised Geminal Power (JAGP).
This wavefunction  has been demonstrated to be effective in 
describing highly correlated diatomic molecules like the $C_2$
as well as $\pi - \pi$ interacting complexes 
\cite{casula2003,casula2004,sorella2007}.
In the present article we present a Variational Monte Carlo (VMC) 
and Lattice Regularized Diffusion Monte Carlo (LRDMC) study 
of the water dimer dissociation energy and dispersion curve,
using as a variational ansatz the JAGP wavefunction.
An important advantage of the JAGP VMC approach resides in the possibility to 
dissect, in a simple way, the energy contributions of the different 
terms composing the wave function, like dynamical electron correlations 
and the intermolecular covalent contribution.
Dynamical electronic correlation associated to the charge fluctuations
and van der Waals interactions, are 
indeed included in the wavefunction within by Jastrow terms, 
whereas static correlation are described by the resonance of  valence
bond singlets in the AGP.
The amount of binding energy arising from the correlated dynamical charge 
fluctuations, related to the vdW forces, can be therefore estimated
by the evaluation of the energy contributions of the Jastrow factors.
On the other side, following the Pauling idea of chemical bonding 
we can calculate the energy contribution of the intermolecular covalent
term and get insight on the 
covalent nature of the hydrogen bonding mechanism. 

%=====================================================================
\section{Computational Methods}\label{method}

{\bf Geometries.}
As nuclear coordinates of the water monomer we used the experimental 
equilibrium geometry \cite{benedict1956} with an O-H bond length of $0.9572 
$ {\AA} and a H-O-H angle of $104.52^{\circ}$. For the dimer we used the 
linear configuration with $C_s$ symmetry, oxygen-oxygen distance
$2.976$  {\AA} \cite{odutala1980,benedek2006} and  $O_1-H_1 
\cdots O_2 $ angle of $180^{\circ}$. We used the internal geometry 
of each monomer as the experimental one.
For the dispersion curve we simply used the geometries 
obtained by shifting away the two monomers along the $O_1-H_1. \cdots O_2$ 
binding axis and keeping fixed their relative orientation. 
Effects of  nuclear relaxation upon binding do not affect our estimations, 
as we verified by calculating the binding energy with the geometry from CCSD(T)
calculations \cite{klopper2000}. 

{\bf Variational Monte Carlo and the JAGP wavefunction.}
As variational ansatz we use the
JAGP wavefunction introduced in references 
\cite{casula2003} and \cite{casula2004}. 
The wavefunction $\Psi_{JAGP}$ of a 
system of $N$ electron is defined by the product of a symmetrical Jastrow 
term $J$ and an antisymmetrical determinantal part $\Psi_{AGP}$: 
\begin{equation} 
\Psi_{JAGP} ( {\bf r}_1,...,{\bf r}_N ) = 
\Psi_{AGP} ( {\bf r}_1,...,{\bf r}_N ) J ( {\bf r}_1,...,{\bf r}_N ) 
\end{equation} 
The determinantal part $\Psi_{AGP}$ is the antisymmetrized product of spin 
singlets. The pairing function in singlet system without spin polarization is 
described by:
\begin{equation}
\Psi_{AGP} = \hat{A} [ \Phi({\bf r}_1^{\uparrow},{\bf r}_1^{\downarrow})\cdots
\Phi({\bf r}_{N/2}^{\uparrow},{\bf r}_{N/2}^{\downarrow})]
\end{equation}
where $ \hat{A} $ is the operator that antisymmetrizes the product of 
$N/2$ geminal singlets 
 $\Phi({\bf r}^{\uparrow},{\bf r}^{\downarrow}) = 
 \psi({\bf r}^{\uparrow},{\bf r}^{\downarrow}) 1/\sqrt{2} 
 ( | \uparrow \downarrow \rangle - | \downarrow \uparrow \rangle) $.
The spatial part of the geminals are expanded over an atomic basis set:
\begin{eqnarray}
\label{geminal}
\psi({\bf r}^{\uparrow},{\bf r}^{\downarrow})&=& \sum_{a,b} \psi_{a,b} ({\bf r}^{\uparrow},{\bf r}^{\downarrow})  \\
 \psi_{a,b} ({\bf r}^{\uparrow},{\bf r}^{\downarrow}) &=&
\sum_{l,m} \lambda_{l,m}^{a,b} \phi_{a,l}({\bf r}^{\uparrow}) \phi_{b,m}
({\bf r}^{\downarrow})
\end{eqnarray}
where the indexes $l,m$ runs over different orbitals centered on nuclei 
$a,b$. 
The Jastrow factor $J$ is further split into one-body,
two-body and three-body terms ($J = J_1 J_2 J_3$). The $J_1$ and $J_2$ 
terms deal with electron-electron and electron-ion correlation respectively. 
The  two body (one body) Jastrow depends only on the relative distance 
$r_{i,j}= |{\bf r}_i - {\bf r}_j |$ between 
each electron pair $(i,j)$ (electron-ion pair) and has been parametrized  
by a simple function $u(r_{i,j}) = ( 1- exp(-b r_{i,j}))/2b$
 that rapidly converges to  a constant 
when $r_{i,j}$  became large.\cite{sorella2007}
In this way the large distance behaviour of the Jastrow is 
determined only by the  
 $J_3$ Jastrow factor, that contains all variational freedom left and 
in particular, as we shall see later on, the slowly decaying vdW correlations.
Therefore we have chosen to parametrize this important part 
of our correlated wave-function in a systematic and exhaustive way, 
similarly to what we have done for the AGP contribution:
\begin{eqnarray} 
\label{3body}
J_3(\textbf{r}_1,...,\textbf{r}_N) &=& 
\exp \left( \sum_{i<j} \Phi^J (\textbf{r}_i,\textbf{r}_j) 
\right)\nonumber \\
\Phi^J&=& \sum_{a,b} \Phi^J_{a,b}  \\
\Phi^J_{a,b} (\textbf{r}_i,\textbf{r}_j) &=& 
\sum_{l,m} g_{l,m}^{a,b}\phi^J_{a,l}(\textbf{r}_i)\phi^J_{b,m} 
(\textbf{r}_j)
\end{eqnarray}
Both the determinantal $\phi_{a,l}$ and and Jastrow $\phi^J_{a,l}$ orbitals 
are expanded on gaussian basis sets centered on the corresponding nuclear 
centers  $a$ and $b$. 
By increasing the atomic basis set one can rapidly reach the 
''complete basis set limit'' because all cusp 
conditions are satisfied by an  appropriate and simple choice of the 
 $J_1$ ( satisfying the electron-ion cusp) and $J_2$ (satisfying the electron-electron cusp) terms.\cite{sorella2007}

All variational parameters, such as the Jastrow parameters, the $\{g\}$ and 
the $\{\lambda\}$ matrices of equations \ref{geminal} and \ref{3body}, 
as well as the exponents and the coefficients of the gaussian orbitals 
 have been optimized by energy minimization following the methods described in refs. 
\cite{sorella2005,umrigar2007}. 

The oxygen valence-core interaction was described using 
the recently reported energy-consistent pseudopotentials \cite{burkatzki2007}.
A VMC calculation for the dimer system with the larger basis set and 
with 0.1mH accuracy run for about twelve hours on eight AMD Opteron 280 CPUs at
the CASPUR computer centre. Full wavefunction optimization was about
a factor 4 more time consuming. 

{\bf Diffusion Monte Carlo.}
A systematic way for improving the quality of a variational wave function is to
perform a Diffusion Monte Carlo calculation, filtering the ground
state properties by a diffusion process \cite{foulkes}.  Actually, due
the presence of the fermionic problem, the DMC is implemented within
the fixed node (FN) approximation \cite{tenhaaf1994}, by imposing that the final
ground state has  the same nodal structure of the trial WF. In this
work we use a slightly modified version of the DMC method, the Lattice
Reguralized Diffusion Monte Carlo (LRDMC).
In this method, the continuum Monte Carlo moves are made 
by discrete finite 
steps defined by two lattice spaces $a$ and $a^\prime$.  
By using an incommensurate ratio $a^\prime/a$ 
the electronic trajectory fills 
the entire space, thus avoiding most lattice artifacts. 
The introduction of the lattice implies that there are 
a finite number of possible moves during the diffusion process,
and this allows one to avoid the locality approximation 
and to restore the upper bound property of DMC.\cite{casula2005,casula2006} 
Within this regularization the exact Hamiltonian
$H$ is  replaced by a lattice reguralized one $H_{a}$ such that
$H_{a} \to H$ for $ a \to 0$ \cite{casula2005}.
 We used as lattice
spaces the values $a=0.1,0.2,0.3,0.5$ a.u. and then the energy were
extrapolated to zero lattice space.

Since the dipole moment operator
does not commute with the Hamiltonian, in LRDMC we evaluated the
estimator $\mu=2\mu_{LRDMC}-\mu_{VMC}$, where $\mu_{LRDMC}$ is the
LRDMC mixed average value extrapolated to $a=0$.
A LRDMC calculation for the dimer system with the larger basis set,
0.1mH accuracy and $a=0.2$, run for about twelve hours 
on eight AMD Opteron 280  CPUs at the CASPUR computer centre.

{\bf DFT Calculations.}
For the sake of comparison we perform DFT calculations 
using a plane wave
basis set as implemented in the CPMD code \cite{car1985}.
For the exchange and the correlation part of the universal functional 
we used BLYP generalized gradient corrections \cite{becke1988,lee1988}
and the hybrid functional B3LYP \cite{becke1993}.
Core electrons were taken into account using norm-conserving 
Troullier-Martins type pseudopotentials \cite{troullier1991}.
We also performed calculations with 
Dispersion-Corrected Atom-Centered Potentials (DCACP)
\cite{vonlilienfeld2004} as described in reference \cite{lin2007}. 
The Kohn-Sham orbitals 
were expanded in plane waves up to a cutoff of 125 Rydberg. 

%=====================================================================
\section{Results and Discussion}
\subsection{Dissociation energy and charge fluctuations}
In this section we report our results on the water dimer at the 
experimental binding distance, 
and we investigate 
the influence of different Jastrow terms of the wavefunction on 
the dissociation energy. In the pairing determinant the oxygen atoms 
are described using a gaussian basis set of $4s4p$ contracted 
to $[1s2p]$, whereas we have only an uncontracted $1s$ shell for the hydrogen.
We verified that the inclusion of a  $d$-wave shell does not affect the binding 
energy giving only a rigid shift of the total energy of the 
dimer and the monomer within LRDMC.

On the contrary, more subtle effects have been observed in the structure of 
the three-body $J_3$ Jastrow factor. This term includes  
in the wave function additional dynamical electron correlations and 
contributes to the proper behavior of the electronic charge 
distribution.
The correct description of the charge correlations reveals crucial for the 
inclusion of the vdW interactions, being 
originated by the correlations between charge fluctuations in different 
spatial regions \cite{dobson2005}.

\begin{table}
\begin{footnotesize}
\begin{center}
\begin{tabular}{|l||c|c|c|c|} \hline\hline
 \multicolumn{4}{|c|}{\bf VMC} \\ \hline
 3-body Jastrow Basis   & $E_{H_2O}$ & $E_{(H_2O)_2}$ & $D_e$ \\ \hline\hline
2s2p-local[O]1s[H] &  -17.2279(1)   &   -34.4585(2) &   -0.0024(4)[-1.5(0.3)] \\ \hline
2s2p[O]1s[H]       &  -17.2388(2)   &   -34.4807(5) &   -0.0031(7)[-1.9(0.4)] \\ \hline
2s4p[O]1s[H]       &  -17.24089(5)  &   -34.4874(1) &   -0.0056(2)[-3.5(0.1)] \\ \hline
2s6p[O]1s[H]       &  -17.24119(8)  &   -34.4886(1) &   -0.0062(4)[-3.9(0.3)] \\ \hline
2s6p[O]1s1p[H]     &  -17.2435(1)   &   -34.4940(1) &   -0.0071(2)[-4.5(0.1)] \\ \hline\hline
 \multicolumn{4}{|c|}{\bf LRDMC} \\ \hline
2s2p-local[O]1s[H] 	 & -17.2576(2) & -34.5228(2) & -0.0076(3) [-4.8(0.2)] \\ \hline
2s2p[O]1s[H] 	         & -17.2613(1) 	 & -34.5303(1) & -0.0077(2) [-4.8(0.2)] \\ \hline
2s4p[O]1s[H] 	         & -17.2619(1) 	 & -34.5315(1) 	 & -0.0077(2) [-4.8(0.2)] \\ \hline
2s6p[O]1s[H] 	         & -17.2619(1) 	 & -34.5314(1) 	 & -0.0076(2) [-4.8(0.1)] \\ \hline
2s6p[O]1s1p[H] 	         & -17.2620(1) 	 & -34.5318(1) & -0.0078(2) [-4.9(0.1)] \\ \hline 
\end{tabular}
\caption{\sf VMC and LRDMC energies for the water monomer and dimer (atomic units).
The dissociation energy calculated as $D_e$=E$_{(H_2O)_2}$-2E$_{H_2O}$ 
is reported in the last column in atomic units and (in square brackets) in 
kcal/mol. }
\label{tab:energy}
\end{center}
\end{footnotesize}
\end{table}

In table \ref{tab:energy} we report the JAGP monomer energy, 
$E_{H_2O}$, the dimer energy, $E_{(H_2O)_2}$, and the dissociation energy
$D_e$ for increasing three-body Jastrow basis sets. The dissociation 
energy of the water dimer has been calculated simply as 
$D_e=E_{dimer}-2E_{monomer}$. 
As the number of $p$-wave shells is increased, we observe an improvement 
of the binding energy, eventually obtaining at VMC level a value of 
$D_e=4.5(0.1)$kcal/mol. 

The reported LRDMC results, extrapolated to the 
$a=0$ limit, appear to have a much faster convergence in term 
of total and dissociation energies. This is due to the nature of the
projection method which accuracy relies only on the nodal surface of the 
trial wave function.
Our results indicate that the VMC optimized nodal surface,
and therefore the corresponding DMC energy, is 
only slightly affected by the basis set extension of
the Jastrow factor.
On the other hand, at the VMC level, the size of the Jastrow factor is 
crucial for improving the binding energy of the dimer. This is probably due
to the role of the Jastrow in localizing charges and in introducing 
dynamical correlations. 
On this regard the role of $p$-wave shells in the binding energy 
will be discussed later in more details.

The LRDMC results are obtained extrapolating at zero lattice space and give 
a binding energy of $ 4.9\pm0.1 $ kcal/mol, in good agreement with the high 
level quantum chemistry calculations:  Klopper {\it et. al.} have
reported $4.99$ kcal/mol and $5.02(5)$ kcal/mol, as basis set limit for MP2
and Coupled Cluster calculations \cite{klopper2000}. 
Recently a QMC study\cite{gurtubay2007} has reported a value of 
$5.4(1)$ kcal/mol, obtained without optimization of the orbitals 
in the determinant, which is directly taken from a  B3LYP calculation.  

Our LRDMC results are 
in the range of previously reported all-electron and pseudopotential QMC 
calculations \cite{benedek2006,gurtubay2007,diedrich2005}. We also agree 
with experimental results although they suffer uncertainty 
due to theoretical estimation of the ZPE.
Actually when compared with the experimental dimer dissociation
energy $D_e^{exp}{}$, the difference between the zero point energy (ZPE)
of the monomer and the dimer should be also taken into account:
$D_e^{exp}=D_0-2{ZPE}^{monomer}+ZPE^{dimer}$. The experimental energy
reported hereafter is therefore corrected by this quantity calculated by
theory or estimated by experiments \cite{benedek2006}.
As pointed out in Ref.\cite{sorella2007} 
the JAGP wavefunction is certainly 
size consistent for the two water monomers, 
only when the complete basis 
set limit is reached for the Jastrow factor.
This property can be used 
to check the basis-set accuracy of the three-body 
Jastrow term. 
To verify the size consistency of the wave function 
we calculated the dissociation energy $D_e^*$ by 
separating the two monomers
at large distance. $D_e^*$ agrees with $D_e$.

\begin{table}
\begin{footnotesize}
\begin{center}
\begin{tabular}{|l||c|c|} \hline\hline
  \multicolumn{3}{|c|}{\bf VMC}\\ \hline
3B Jastrow Basis           & $\mu_{H_2O}$ [Debye]& $\mu_{2[H_2O]}$ [Debye]  \\ \hline\hline
2s2p-local[O]1s[H] &  2.116(17) & 2.805(20)      \\ \hline
2s2p[O]1s[H] 	   &  1.935(12) & 2.834(23)      \\ \hline
2s6p[O]1s[H] 	   &  1.880(8)   & 2.692(14)     \\ \hline
2s6p[O]1s1p[H]     &  1.890(8) 	& 2.597(12)    \\ \hline\hline  
\multicolumn{3}{|c|}{\bf Extrapolated } \\ \hline
2s6p[O]1s[H] 	   &  1.874(10)   & 2.648(18)     \\ \hline
2s6p[O]1s1p[H]     &  1.870(10)   & 2.603(13)    \\ \hline  
\end{tabular}
\caption{\sf Dipole moment of water monomer and dimer. VMC estimates are reported in the top part of the table. 
Extrapolated values are reported in the bottom part of the table.}
\label{tab:dipole}
\end{center}
\end{footnotesize}
\end{table}

In QMC calculations, correlation functions  different from the energy are often 
very sensitive to the quality of a wavefunction. 
We have therefore calculated the monomer and dimer dipole moment 
$\mu$ that can be easily computed at the VMC level and at LRDMC level 
using the mixed estimator.
In the case of the water monomer both the variational and the
LRDMC dipole moments are rather close to the experimental
value of $1.855 D$ \cite{franks1972}.
The correction introduced by LRDMC is a slight downshift
of the extrapolated estimator, $\mu = 1.870(10)$ $D$, in 
agreement  with other QMC calculations \cite{gurtubay2007} 
and {\it ab initio} methods \cite{coutinho2003}.

We turn now the attention to the role of dynamical correlations included 
through the Jastrow term.
As discussed above the inclusion of  $p$-wave orbitals 
in the $J_3$ Jastrow term has a significant effect on 
the binding energy. Similarly, the effect of the $J_3$ basis set
is also visible on the dipole moments in table \ref{tab:dipole}.
One reason of this influence can be attributed to electrostatic 
interactions, since the 3-body term is important for the charge
distribution. Another relevant effect of the $J_3$ term is
the modulation of the van der Waals interactions.
Van der Waals forces are indeed
quantum mechanical effects originating by the interaction between 
instantaneous dipoles, or, using a second order perturbation
theory perspective \cite{cohen1977}, 
by the correlated transition of a couple of electrons from
occupied to unoccupied states.
Given thus two atomic centers $a$ and $b$ at large 
distance the $J_3$ term in Eq. \ref{3body} 
can be expanded for small value of $g_{ij}^{a,b}$ 
and then applied to a  single  geminal pair (see Eq.\ref{geminal}),  
$\psi_{a,b} (r^{\uparrow},r^{\downarrow})$.
The result can be viewed as  a correlated transition of two  electrons 
 located in different atomic centers 
 from 
occupied  orbitals    to unoccupied  orbitals with higher 
angular momentum\cite{cohen1977}.
 More generally the  effect of the $J_3$ 
at large distance has the same  structure of the vdW perturbative term
if on each atomic center the basis used for the Jastrow contains odd orbitals 
with respect to the spatial reflection, namely  
when the Jastrow basis set contains at least $p$ wave orbitals.
In principle a small vdW contribution can derive also from high angular 
momentum orbitals included in  the geminal expansion. 
In this work however, in order to disentangle the   genuine vdW  contribution, 
we have avoided to use polarization orbitals in the AGP, that, as discussed 
before,  do not affect the binding energy.
In this way
the  instantaneous correlated polarization induced by the $J_3$ term allows 
to include vdW interactions in a transparent variational form.

To understand the effects of the  $J_3$ terms on the 
dissociation energy, we calculated 
the variational energy of the wavefunction obtained excluding inter-molecular 
$g_{l,m}^{a,b}$ terms in Eq. \ref{3body} as reported in table
\ref{tab:vdw}.
In particular we considered the $H-O$ and $O-O$ contributions in
the $p$-$p$ channel 
and eventually we eliminated all intermolecular terms (last row of the 
table \ref{tab:vdw}).
Data show non-additivity of the energy loss, as expected by 
interactions arising from polarization effects \cite{landau1981}.
Among the $p-p$ wave contribution, the oxygen-oxygen channel 
seems to be the most relevant term in the 
Jastrow expansion. It is worth noting that the total dipole 
moment of the dimer depends only weakly on the intermolecular 
$J_3$ Jastrow terms, see table \ref{tab:vdw}. 
This indicates that the distribution of the electronic charge 
is not much affected by the missing terms. The energy differences
are then due to the part of the dynamical correlation 
involving correlated excitations to $p$ states.
The energy loss in the binding energy can therefore be attributed within our
formalism to van der Waals interactions.

It is of interest to compare our result to previous calculations based on 
symmetry-adapted-perturbation-theory (SAPT) \cite{mas1997,misquitta2005} 
that estimated the contribution of dispersion forces to the water dimer 
hydrogen bond. 
This contribution amounts to about -1.75 kcal/mol as reported in
table V of Ref. \cite{misquitta2005}. Albeit 
the energy is not partitioned the same way in the two approaches,
the assessment given by SAPT in good agreement with our 
estimation of -1.5(2) kcal/mol. 

\begin{table}
\begin{footnotesize}
\begin{center}
\begin{tabular}{|l||c|c|c|c|} \hline\hline
Pairing terms in $J_3$ 
& E$_{(H_2O)_2}$ (a.u) & $\Delta E$ (kcal/mol) & 
 $\mu_{(H_2O)_2}$ [D]  \\ \hline\hline
 Full $\{g_{l,m}^{a,b}\}$ matrix  &  -34.4940(1) & 0.0 & 2.597(12) \\ \hline
 (p[H])$_1$ (p[H])$_2$    &  -34.49372(9) & +0.2(1) & 2.621(12) \\ \hline
 (p[O])$_1$ (p[H])$_2$    &  -34.4938(1) & +0.1(1) &  2.623(12) \\ \hline
 (p[H])$_1$ (p[O])$_2$    &  -34.4935(1) & +0.3(1) &  2.610(13) \\ \hline
 (p[O])$_1$ (p[O])$_2$    &  -34.4918(1) & +1.4(1) &  2.628(12) \\ \hline\hline
% (s[H])$_1$ (p[O])$_2$    &  -34.4915(1) \\ \hline
 intermolecular p-p $g_{l,m}^{a,b} = 0$ & -34.4916(3)   & +1.5(2) & 2.637(13) \\ \hline\hline
\end{tabular}
\caption{\sf VMC energy and dipole moment of the water dimer
for different Jastrow $J_3$ terms.
The different $J_3$ are obained by canceling the $p-p$ electronic 
correlation between atomic centers belonging to different molecules.
The atomic center of the $p$ wave is indicated between square 
brackets and the water molecule index is indicated by the pedex (1 or 2).
The energy difference $\Delta E$ with respect the complete 
$g_{l,m}^{a,b}$ matrix, first line, is also reported in kcal/mol.}
\label{tab:vdw}
\end{center}
\end{footnotesize}
\end{table}

%#####################################################################

%=====================================================================
\subsection{Dispersion curve}

\begin{figure} 
%\begin{center} 
\includegraphics[angle=-90,width=0.50\textwidth]{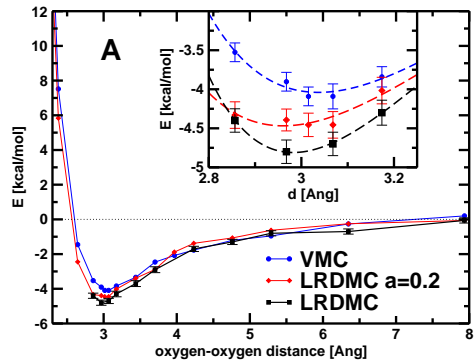}
\includegraphics[angle=-90,width=0.50\textwidth]{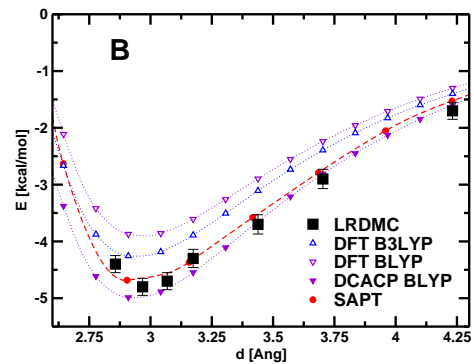}
\caption{(Color on line)\sf Water dimer dissociation. The total energy of the 
water dimer is reported as a function of the 
oxygen-oxygen distance. In Panel A the VMC and LRDMC results are reported. 
In the inset graph the behavior around the minimum is zoomed in. 
In Panel B the LRDMC curve is compared with other methods. 
SAPT values have been taken from ref. \cite{wu2001}. }
%\end{center} 
\label{fig:dispersion}
\end{figure}

The VMC and LRDMC dispersion curve of the water dimer is reported in 
Fig.\ref{fig:dispersion}A.
It has been calculated by computing
the total dimer energy as a function of the oxygen-oxygen distance without 
changing the internal geometry and the relative orientation of the monomers. 
We used the 2s6p[O]1s[H] basis set for the 3-body Jastrow, which, 
as reported before, guarantees size consistency during the 
dissociation process at large distances.

The attractive tail of the water-water interaction potential 
is dominated by a dipole-dipole interaction energy.
A polynomial fit for $d \ge 3.5$ {\AA} shows that $E_{2(H_2O)} 
\sim d^{\alpha}$ with  $\alpha =3.2-3.3$ for the VMC and the extrapolated
LRDMC curves, respectively.

\begin{figure} 
\begin{center} 
\includegraphics[angle=0,width=0.50\textwidth]{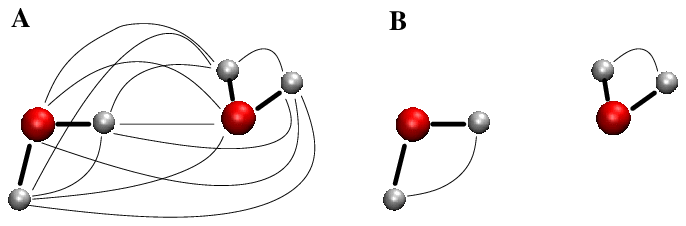}
\includegraphics[angle=-90,width=0.50\textwidth]{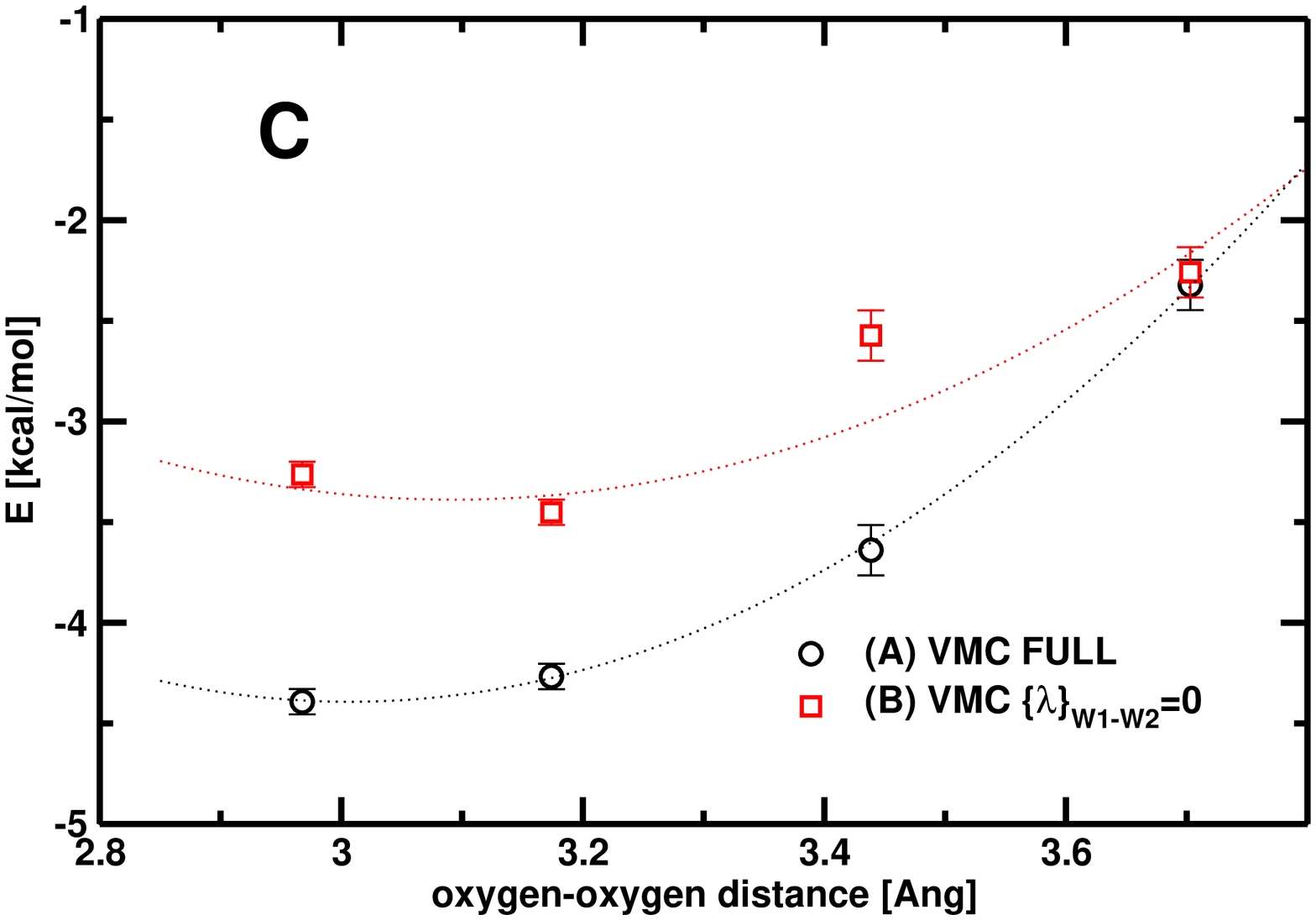}
\caption{(Color on line)\sf Top panel: Pictorial view of the intermolecular pairing term of the determinant part of the wavefunction. In Panel A all the intermolecular pairing are drawn. In Panel B we represent the wavefunction with the
intermolecular pairing term set to zero. In the bottom part of the Figure
the total energy of the 
 water dimer is reported as a function of the 
oxygen-oxygen distance between the two monomers. Two wavefunctions are compared, one with all the intermolecular pairing term optimized (black circle) and the 
other one with  intermolecular pairing terms set to zero (red square).}
\label{fig:covalent}
\end{center} 
\end{figure}
The behavior of the dispersion curve at short distances is shown in the inset 
of Fig. \ref{fig:dispersion}A, together with a fit performed using a 
Morse potential.  
At the VMC level, the minimum of the curve as obtained by the fitting 
procedure, is at distance $d=3.037(4)$ {\AA}, which is slightly shifted with 
respect to $d=2.976$ {\AA}  reported by experiments \cite{odutala1980}. 
However, it should be noted that, considering the error bars, 
the curve results to be rather flat around the equilibrium distance. 
LRDMC with a lattice space $a=0.2$ a.u., and the LRDMC  
extrapolation to zero lattice space $a\rightarrow 0$, improve the location 
of the equilibrium distance. In this latter case 
the fitted minimum is at $d=2.982(1)$ {\AA}, which is very close to
the experimental value.
In figure \ref{fig:dispersion}B we report a comparison with 
pure or empirically 
parametrized Density Functional methods and symmetry-adapted 
perturbation theory (SAPT)\cite{mas1997,wu2001}. 
Data show that pure BLYP and B3LYP curves underestimate the 
dissociation energy whereas 
calculations performed with empirically parametrized 
DCACP pseudopotentials are closer to our LRDMC curve. SAPT curve  
is on the top of our results with small discrepancies at 
very short distance.

\subsection {Covalent contribution to hydrogen bonding}

In the proximity to the equilibrium distance 
the interplay between electrostatic and pure quantum effects
is expected to be relevant. 
Although a unique and commonly accepted definition of covalent contribution 
in hydrogen bond is still missing, 
within the formalism of the JAGP wavefunction,
we can define the covalent energy contribution as the
energy contribution arising from the 
intermolecular pairing terms of the RVB determinantal part of the wavefunction
(see Eq. \ref{geminal}).

The ``chemical bond'' between the two molecules is indeed due to the 
superposition of all singlet terms in the geminal expansion that 
connect two nuclei belonging to different water molecules. 
This is schematically illustrated in the panel A of Fig. \ref{fig:covalent}.
In order to evaluate the covalent contribution we proceed as follows.
We cut  
the intermolecular pairing valence bonds in the pairing 
function,  by imposing $\psi_{a,b}=0$ if $a$ and $b$ 
belong to different 
monomers (as sketched in panel B of Fig.\ref{fig:covalent}). 
Then the wave function is re-optimized  with the above constraint 
in order to  
correctly include the electrostatic effects and the slowly 
decaying vdW correlations present in our Jastrow factor. 
In Fig. \ref{fig:covalent} C we report, as a function of 
the oxygen-oxygen distance, the binding energy 
calculated with the full wavefunction (circle)
and with the wave function lacking the
intermolecular valence bond terms (square).
The difference between the two curves vanishes
as the molecules reach a
O-O distance of $3.5 $ {\AA}. 
We point out that, by  cutting  the intermolecular pairing terms,  
the minimum of the  
energy dispersion slightly shifts to a larger equilibrium distance.

At the equilibrium distance we found that the contribution of the 
intermolecular pairing terms, computed at the VMC level, 
is $\Delta_{inter} =1.1(0.1) kcal/mol$, corresponding to 
about $24\%$ of the computed dimer binding energy. 
Our estimate of the covalent contribution, defined above,
can be compared  to 
what other authors found using different theoretical frameworks and 
different definition and that is generally referred to as intermolecular 
charge transfer (CT) \cite{morokuma1977,kollman1985,mo2000,iwata2007}.
%here
In the seminal works based on Morokuma decomposition of the 
binding energy \cite{morokuma1977,kollman1985} 
the CT contribution to hydrogen bonding
is estimated in the range $-1.3 \div -1.8$ kcal/mol., 
thus about 25\% of the total binding energy.
A slight smaller contribution, about $11\%$  resort from block-localized 
wave function approach proposed by Mo {\it et al.} \cite{mo2000}. They also
reported that CT contribution vanishes at about $3.5$ {\AA} in good 
agreement with our finding.  
%here
It is interesting to observe that the damping of the oscillation at
$d \sim 4$ {\AA} of the Fourier Transforms of the Compton profile, has 
been interpreted in ref. \cite{barbiellini2002} as a cut-off for the covalent
contribution. 
\cite{barbiellini1999,barbiellini2002}. However such an interpretation
of the experiments is not fully accepted \cite{ghanty2000,romero2000}.

%=====================================================================
\section{Conclusions}

The understanding of hydrogen bond systems is still a challenge
for computational chemistry. Even for small molecular systems
the weakness of the interactions and the critical role of electron 
correlation require to use affordable correlated quantum 
chemistry methods.
The interplay between interactions different in nature, 
such as dispersion forces and intermolecular charge transfer are
in many cases crucial for the proper description 
of the bond properties. 

We have shown that the Quantum Monte Carlo method is effective for describing
the hydrogen bond between two water molecules. The calculated 
binding energy matches the experimental value and the estimates from other
advanced methodologies. Good agreement with experiments
is also achieved for the computed dipole moments. 
Thanks to the good size scaling properties and the embarrassingly parallelism 
of QMC algorithms, these methods appear extremely competitive 
in the context of massive parallel computation.

Moreover, some conceptual advantages rely on the structure of the 
AGP wavefunction, a correlated valence bond representation of the electronic 
system. The AGP formalism 
gives the possibility to work back on an  intuitive picture 
of localized 
chemical bonds  such as the Pauling's superposition of Lewis structures.
Thanks to the fully correlated structure of the 
wavefunction, this picture can be used 
without compromises in term of accuracy.

Upon interpretation of the wavefunction terms, we  
estimate at the VMC level the covalent contribution to account for 
1.1(2)  kcal/mol. A similar contribution to the binding 
energy is given by correlated dipolar vdW fluctuations that 
account for 1.5(2) kcal/mol. 

The quality of our results on the water dimer encourages the application of the
method to larger hydrogen bonded systems such as water clusters 
or small biomolecules.
To reduce the computational costs it would be desirable to keep 
down the number of variational parameters
when the size of the system increase. In this respect, different 
strategies are under investigation. 

%=====================================================================
\section{Acknowledgments}
F.S, L.F., and L.G. acknowledge CASPUR for  
large computational resources and technical and financial support. 
We thanks C. Filippi for providing us with the 
oxygen pseudopotential before publication.
L.G. thanks the Centro Fermi of Rome and for computational resources.
L.S.,S.S. and L.G. are gratefull to INFM and Cineca for the grant n. 806 "Progetti Calcolo 2007". 
%Authors thanks G.B. Bachelet, B. Barbiellini, G. Galli,
%S. Moroni, C. Pierleoni, and U. Rothlisberger for
%useful discussions and comments on the manuscript.

%\clearpage
%\bibliographystyle{achemso}
%\bibliography{bibliografia}
\providecommand{\url}[1]{\texttt{#1}}
\providecommand{\refin}[1]{\\ \textbf{Referenced in:} #1}

\end{document}